\documentclass[journal]{IEEEtran}
\usepackage{graphicx}
\usepackage{epstopdf}
\usepackage{multirow}
\usepackage{amssymb}
\usepackage{amsmath}
\usepackage[linesnumbered,algoruled,boxed,lined]{algorithm2e}
\usepackage{color}
\usepackage{amssymb}
\usepackage{bm}
\usepackage{booktabs}
\usepackage{ntheorem}

\theoremseparator{.}
\newskip\theorempreskipamount
\newskip\theorempostskipamount

\theorembodyfont{}

\allowdisplaybreaks[4]

\ifCLASSOPTIONcompsoc
  \usepackage[nocompress]{cite}
\else
  \usepackage{cite}
\fi

\ifCLASSINFOpdf
\else
\fi
\begin{document}
%
\title{Securing the Sky: Integrated Satellite-UAV Physical Layer Security for Low-Altitude Wireless Networks}
%
%
%
%

\author{
Jiahui Li,
Geng Sun,~\IEEEmembership{Senior Member,~IEEE},
Xiaoyu Sun,
Fang Mei, 
Jingjing Wang,~\IEEEmembership{Senior Member,~IEEE},\\
Xiangwang Hou, 
Daxin Tian,~\IEEEmembership{Fellow,~IEEE},
and Victor C. M. Leung,~\IEEEmembership{Life Fellow,~IEEE}

	
	\thanks{

        \par Jiahui Li, Xiaoyu Sun, and Fang Mei are with the College of Computer Science and Technology, Jilin University, Changchun 130012, China (e-mails: lijiahui@jlu.edu.cn; xiaoyus23@mails.jlu.edu.cn; meifang@jlu.edu.cn). 

        \par Geng Sun is with the College of Computer Science and Technology, Jilin University, Changchun 130012, China, and with Key Laboratory of Symbolic Computation and Knowledge Engineering of Ministry of Education, Jilin University, Changchun 130012, China; he is also affiliated with the College of Computing and Data Science, Nanyang Technological University, Singapore 639798 (e-mail: sungeng@jlu.edu.cn).

        \par Jingjing Wang is with the School of Cyber Science and Technology, Beihang University, Beijing, 100191, China. (e-mail: drwangjj@buaa.edu.cn).

        \par Xiangwang Hou is with the Department of Electronic Engineering, Tsinghua University, Beijing, 100084, China. (e-mail: xiangwanghou@163.com).

        \par Daxin Tian is with the State Key Laboratory of Intelligent Transportation System, Beijing Key Laboratory for Cooperative Vehicle Infrastructure Systems and Safety Control, School of Transportation Science and Engineering, Beihang University, Beijing 100191, China, and also with the Zhongguancun Laboratory, Beijing 100081, China (e-mail: dtian@buaa.edu.cn).

        \par Victor C. M. Leung is with the College of Computer Science and Software Engineering, Shenzhen University, Shenzhen 518060, China, and also with the Department of Electrical and Computer Engineering, The University of British Columbia, Vancouver V6T 1Z4, Canada (e-mail: vleung@ieee.org).
        
        \par \textit{Corresponding authors: Geng Sun; Fang Mei.} 
        \protect
	}
}
\IEEEtitleabstractindextext{%

\begin{abstract}
Low-altitude wireless networks (LAWNs) have garnered significant attention in the forthcoming 6G networks. In LAWNs, satellites with wide coverage and unmanned aerial vehicles (UAVs) with flexible mobility can complement each other to form integrated satellite-UAV networks, providing ubiquitous and high-speed connectivity for low-altitude operations. However, the higher line-of-sight probability in low-altitude airspace increases transmission security concerns. In this work, we present a collaborative beamforming-based physical layer security scheme for LAWNs. We introduce the fundamental aspects of integrated satellite-UAV networks, physical layer security, UAV swarms, and collaborative beamforming for LAWN applications. Following this, we highlight several opportunities for collaborative UAV swarm secure applications enabled by satellite networks, including achieving physical layer security in scenarios involving data dissemination, data relay, eavesdropper collusion, and imperfect eavesdropper information. Next, we detail two case studies: a secure relay system and a two-way aerial secure communication framework specifically designed for LAWN environments. Simulation results demonstrate that these physical layer security schemes are effective and beneficial for secure low-altitude wireless communications. A short practicality analysis shows that the proposed method is applicable to LAWN scenarios. Finally, we discuss current challenges and future research directions for enhancing security in LAWNs.
\end{abstract}
}

\maketitle

\IEEEdisplaynontitleabstractindextext

%
\IEEEpeerreviewmaketitle

%
\section{Introduction}
\label{sec:intro}

\par The emerging low-altitude economy (LAE) represents a significant shift in how we utilize airspace up to 3,000 meters above ground level~\cite{Wang2025}. Specifically, LAE encompasses various activities, including cargo transportation, low-altitude inspections, and passenger services conducted using both manned and unmanned aerial vehicles (UAVs) such as drones and electric vertical take-off and landing aircraft (eVTOL). To support this burgeoning sector, low-altitude wireless networks (LAWNs) have been proposed as a critical enabling infrastructure. In particular, LAWNs integrate various non-terrestrial network (NTN) components, particularly satellite and UAV platforms, to establish robust communication frameworks that extend beyond traditional terrestrial infrastructure~\cite{He2025}. These integrated satellite-UAV networks have garnered significant attention in the development of forthcoming 6G networks due to their advantages in providing wider coverage, higher mobility, and increased resilience to disruptions in challenging environments.

\par Different from traditional UAV deployments, low-altitude wireless networks (LAWNs) involve numerous flying vehicles that both function as base stations and provide specific services like air taxis. However, data security poses a critical challenge in establishing LAWNs due to the high mobility of aerial vehicles and weather challenges. Conventional encryption and decryption protocols may not be applicable to UAV-based LAWNs due to their massive computational resource requirements. In such scenarios, physical layer security emerges as a promising solution to ensure information security in wireless communications~\cite{Lv2024}. This approach capitalizes on the distinctive characteristics of wireless channels to conceal legitimate information from potential eavesdroppers, thereby reducing the computational overhead associated with encryption and decryption operations.

\par In the LAWN environment, UAV swarms are often deployed to provide extensive network coverage. As such, collaborative beamforming stands out among various physical layer security schemes for UAV swarms as it may enhance transmission capacity and energy efficiency simultaneously. Specifically, the antenna of each UAV can be considered to be an independent element of a virtual antenna array. Similar to a co-located antenna array, the beamforming coefficients of these antennas can be carefully designed to control the signal distribution of the antenna system. Furthermore, the positions of the UAVs can be rapidly fine-tuned to alter the antenna aperture and shape, optimizing the beam pattern. Once optimized, the signal strength towards unauthorized devices can be suppressed, thereby achieving physical layer security without massive trajectory designs.

\par Some existing studies have provided enlightening sights in UAV-enabled collaborative beamforming for physical layer security in LAWNs. For instance, the authors in~\cite{Huang2025} proposed a dual UAV cluster-assisted system via collaborative beamforming to achieve physical-layer security in maritime wireless communications, where one UAV cluster forms a maritime UAV-enabled virtual antenna array relay to forward data signals to the legitimate vessel, while another UAV cluster functions as a jammer to send interference signals to potential eavesdroppers. Likewise, the authors in~\cite{Liu2024} investigated a UAV-assistant air-to-ground communication system with multiple UAVs forming a UAV-enabled virtual antenna array, proposing a multi-objective optimization approach to simultaneously maximize the transmission rate and minimize energy consumption by optimizing UAV positions and excitation current weights. Additionally, in~\cite{Li2024}, the authors introduced distributed collaborative beamforming into UAV swarms to counter eavesdropper collusion in aerial communications. By constructing two UAV swarms as virtual antenna arrays for two-way communication, the authors minimized both the two-way known secrecy capacity and maximum sidelobe level while also reducing energy consumption.

%
\begin{figure*}
  \centering
  \includegraphics[width=6.9 in]{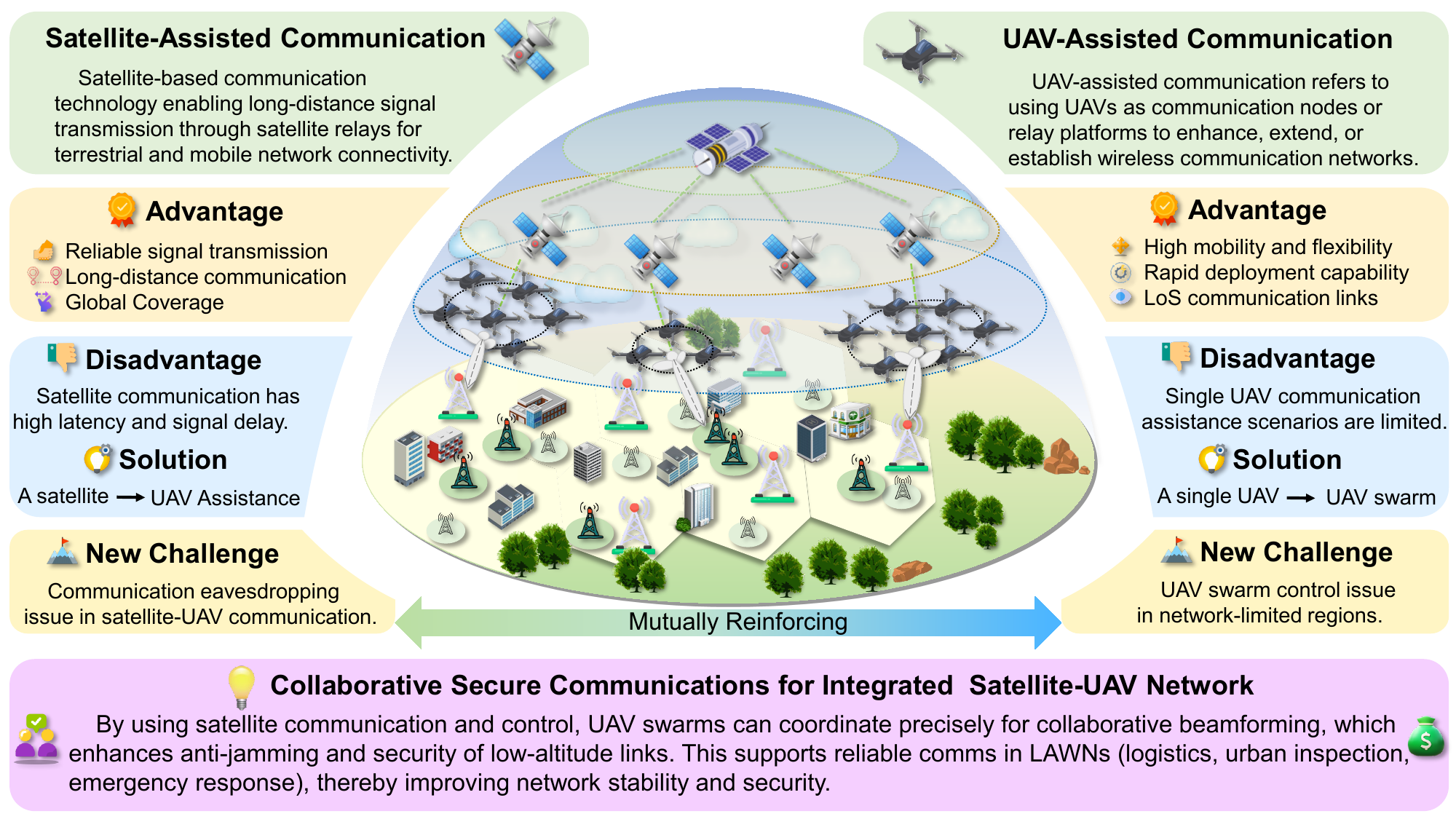}
  \caption{Structure of the integrated satellite-UAV secure communication paradigm for LAWN applications. This paradigm enhances physical layer security through collaborative beamforming, where UAV swarms provide low-latency, high-throughput secure connections while satellite networks handle control information, particularly suitable for regions with inadequate network infrastructure.}
  \label{fig:structure}
\end{figure*}

\par However, in regions with inadequate network infrastructure, controlling and coordinating UAV swarms pose significant challenges to the existing works which require precise control of the UAV swarm. Thus, as shown in Fig.~\ref{fig:structure}, we aim to propose a novel integrated satellite-UAV secure communication paradigm specifically designed for LAWN applications in this work. We introduce a collaborative beamforming-based physical layer security scheme where UAV swarms provide low-latency and high-throughput secure connections while satellite networks handle the controlling information. We begin by presenting some preliminaries, including integrated satellite-UAV networks, physical layer security, UAV swarms, and collaborative beamforming for LAWN applications. Then, we introduce several promising typical integrated satellite-UAV secure communications applications in the LAWNs. Following this, we present two case studies that demonstrate the benefits and insights of our considered systems in realistic LAWN scenarios, and conduct a practicality analysis. Finally, we discuss current issues and future directions for implementing secure low-altitude wireless networks.

%
\section{Preliminaries}
\label{sec:preliminaries}

\par In our integrated satellite-UAV secure systems for LAWN applications, controlling and scene information are crucial for achieving physical layer security. We provide a brief overview of the essential preliminaries as follows.

%
\subsection{Integrated Satellite-UAV Networks for LAWNs}
\label{ssec:SN}

\par Satellite networks utilize satellites in Earth's orbit to enable wireless communication over vast geographic areas~\cite{Wu2025}. Specifically, the satellites in these networks are equipped with sophisticated communications systems, including antennas, transponders, and processors, enabling them to receive and transmit signals from terrestrial or airborne systems. In LAWNs, satellite networks are particularly valuable for supporting low-altitude operations across extensive geographical areas, including remote or rural locations, open oceans, and airspace where traditional terrestrial networks may be unavailable or inadequate~\cite{Ozturk2025}.

\par However, satellite networks also have some technical limitations, such as the potential signal delay due to the distance between satellites and ground-based systems, susceptibility to interference, and the need for precise orbit and frequency coordination. In LAWN applications, we can reduce the direct communications between satellites and terrestrial devices. Instead, we can leverage the robust remote sensing capabilities and extensive coverage of satellites to control and deploy UAV networks in network-limited areas of the low-altitude airspace. As such, the advantages of UAV and satellite networks are integrated to achieve low-latency and large-bandwidth network coverage for diverse applications such as cargo delivery, aerial surveillance, and air taxi services. 

%
\subsection{Physical Layer Security for LAWNs}
\label{ssec:PLS}

\par Physical layer security serves as a viable approach to ensure information-theoretic level security under the threat of eavesdroppers in LAWN environments. Specifically, the signals directed towards eavesdroppers are intentionally weakened, creating confusion and hindering their ability to intercept secure communications. In general, the secrecy capacity, also known as secrecy rate, can be defined as $\max \{ R_l, R_e \}$, where $R_l$ and $R_e$ are the achievable rates of the legitimate receiver and eavesdropper, respectively~\cite{Cai2025}. In traditional terrestrial wireless networks, power allocation or beamforming, which require channel state information (CSI) or partial CSI of the eavesdropper, are commonly employed for realizing physical layer security.

\par Regarding satellite-assisted UAV networks in LAWN scenarios, physical layer security presents both opportunities and challenges. On the one hand, due to the high LoS probability in low-altitude airspace, aerial networks are more vulnerable to eavesdropping, as eavesdroppers can exist in three-dimensional space and receive stronger signals, making it challenging to achieve physical layer security. On the other hand, UAVs and satellites are equipped with remote sensing capabilities, allowing them to detect potential eavesdroppers in the low-altitude airspace and obtain partial CSI via visual and signal detection~\cite{Fontanesi2025}, which may support physical layer security communications. Moreover, UAVs can be controlled by satellite networks to reach remote and inaccessible areas. Hence, there is a need for novel methods and architectures that can provide physical layer security solutions in such emerging network paradigms.

%
\subsection{UAV Swarm for LAWN Applications}
\label{ssec:UAV_Swarm}

\par In LAWN applications such as industrial or agricultural processes, large-scale Internet-of-things networks, cargo delivery, or air taxi services, a single UAV cannot support various requirements such as monitoring, sensing, or assisting terrestrial networks~\cite{Adil2024}. In such scenarios, utilizing a UAV swarm or a group of coordinated UAVs can offer enhanced communication capabilities and efficiency for low-altitude operations. For instance, different UAVs can cover different areas simultaneously, allowing for broader coverage and more efficient data collection and communication tasks in the low-altitude airspace.

\par However, in network-limited regions of LAWNs, controlling and coordinating UAV swarms pose significant challenges. UAV swarms often have limited transmission ability, making it difficult to establish a reliable connection with remote base stations. Interacting with a UAV swarm to design a reasonable deployment scheme based on the dynamic environment, especially for precise control such as enabling collaborative beamforming, can be challenging. Therefore, it is essential to conduct research and investigation into novel UAV swarm-controlling paradigms that enable intelligent UAV swarm deployment for secure LAWN applications.

%
%
\subsection{Collaborative Beamforming for LAWN Security}
\label{ssec:CB}

\par In the preparation process of collaborative beamforming for LAWN applications, each antenna of an independent communication device is abstracted as an array element~\cite{Li2024a}. Then, these array elements will synchronize their time, phase, and frequency to form a virtual antenna array. In the transmission process, the array elements first aggregate or share data with each other. Then, similar to co-located antenna arrays, the array elements of the virtual antenna array send signals to the receiver simultaneously, thereby generating a gain in the received signal-to-noise ratio. 

\par The signals generated by the virtual antenna array are distributed as a beam pattern. Specifically, the lobe towards the targeted receiver is defined as the mainlobe, while others are sidelobes. In LAWN applications, we can improve the gain of the mainlobe and suppress the strengths of sidelobes for achieving physical layer security. In particular, we can carefully control the excitation current weights of the array elements and properly deploy their positions to minimize the sidelobes towards the detected eavesdroppers~\cite{Li2024a}, thereby purposefully enhancing the secrecy capacity in low-altitude wireless networks.

%
\section{Opportunities for Integrated Satellite-UAV Secure Communication in LAWNs}
\label{sec:typical_applications}

%
\begin{figure*}
  \centering
  \includegraphics[width=6.9 in]{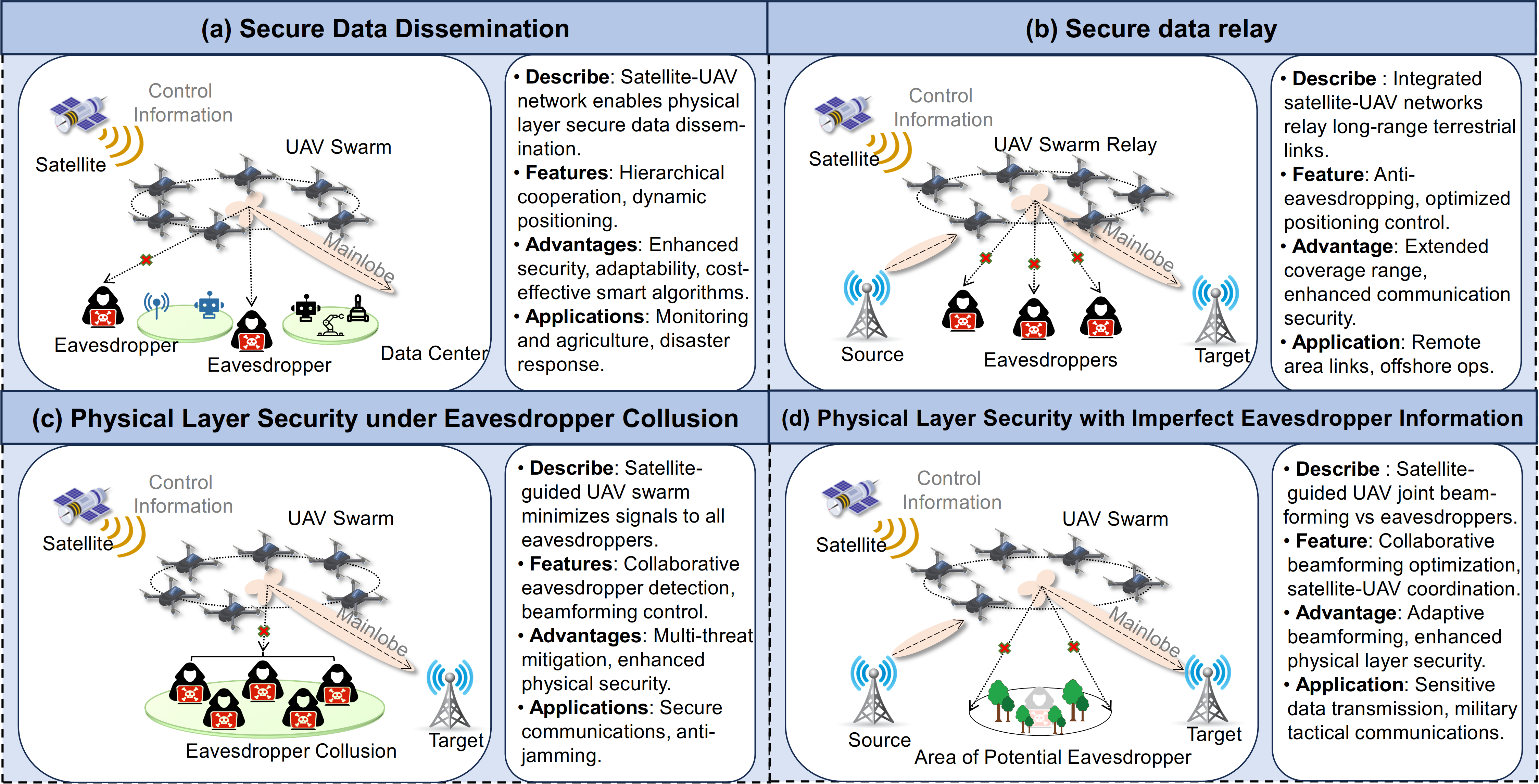}
  \caption{Opportunities for integrated satellite-UAV secure communication in LAWNs. (a) Secure data dissemination for LAWN services; (b) Secure data relay for LAWN infrastructure; (c) Physical layer security under eavesdropper collusion in LAWNs; (d) Physical layer security with imperfect eavesdropper information in LAWNs.}
  \label{fig:typical_application}
\end{figure*}

\par Based on the preceding presentations, it is evident that UAV swarms can leverage satellite assistance to perform collaborative beamforming for communication missions and physical layer security in infrastructureless areas of LAWNs. This allows UAVs to achieve effective deployment schemes and accurate remote sensing information from satellites, while satellites can deploy UAVs to implement low-latency, high-bandwidth security strategies. 

\par However, this secrecy method is still in its infancy, and existing works on UAV collaborative secure schemes only consider UAVs to be able to obtain complete environment information. For instance, in \cite{Li2024a}, the authors proposed a multi-objective optimization method to achieve physical layer security based on perfect channel state information (CSI) and UAV cooperation. In reality, when the UAV swarm performs various LAWN communication tasks with specific secure transmission requirements, CSI, eavesdropper location, and other environmental information are hard to be provided by basic infrastructure. In the following, we will introduce some typical applications of UAV collaborative secure schemes in LAWNs and show how the satellite networks assist the UAV swarm.

%
\subsection{Secure Data Dissemination for LAWN Services}
\label{ssec:application1}

\par In various LAWN applications such as industrial inspection, agriculture monitoring, or cargo tracking, a set of UAVs is often deployed to perform surveillance and sensing tasks in remote regions. These UAVs need to periodically transmit the acquired data to data fusion centers. In this scenario, as illustrated in Fig.~\ref{fig:typical_application}(a), an integrated satellite-UAV network can provide a physical layer secure data dissemination process. Specifically, satellites and UAVs jointly detect terrestrial information via camera or signal radar in the low-altitude airspace. Then, the satellite forwards the environment to the data fusion center for designing a deployment scheme. Afterward, satellites issue the deployment scheme to deploy the positions and excitation current weights of the UAVs, guiding a null towards the eavesdropper direction to secure the LAWN communications.

%
\subsection{Secure Data Relay for LAWN Infrastructure}
\label{ssec:application2}

\par Integrated satellite-UAV networks can also serve as data relays for two-way communications between two long-range terrestrial base stations in the LAWN environment. Due to the higher LoS probability in low-altitude airspace, the transmission efficiency of the relay can be significantly improved compared to the conventional terrestrial relay. However, UAV swarms acting as relays are also more vulnerable to eavesdropping threats and harder to control in network-limited regions of the low-altitude airspace. In this case, as depicted in Fig.~\ref{fig:typical_application}(b), satellites can control the UAV swarm to fly to the midpoint between two remote base stations and provide remote sensing terrestrial information and deployment schemes to the UAV swarm. Subsequently, satellites control both the UAV swarm and the two base stations to perform the relaying tasks. By utilizing optimized positions and excitation current weights, the UAV swarm can serve different terrestrial devices while minimizing the signal obtained by eavesdroppers in the low-altitude airspace.

%
\subsection{Physical Layer Security under Eavesdropper Collusion in LAWNs}
\label{ssec:application3}

\par Eavesdroppers in LAWN environments may collude based on signal detection (\textit{e.g.}, utilizing maximum ratio combining technique), which leads to the worst wiretap case. Moreover, the distribution of multiple eavesdroppers in low-altitude airspace may be random and dispersed, implying that the physical layer security achieved by trajectory design is difficult to realize. In this case, as shown in Fig.~\ref{fig:typical_application}(c), satellites can forward the potential eavesdropper locations detected by remote sensing and UAV radars~\cite{Khawaja2025} to the edge servers for designing the deployment scheme. Then, satellites control the UAV swarm to form an intelligent virtual antenna array to minimize the signals of all eavesdroppers by controlling the beam pattern in the LAWN communication network. 

%
\subsection{Physical Layer Security with Imperfect Eavesdropper Information in LAWNs}
\label{ssec:application4}

\par The positions of eavesdroppers in LAWN environments can be detected through remote sensing using satellites, as well as cameras and radars on UAVs. However, due to the limited performance of these devices, there may be errors in the detection accuracy~\cite{Khawaja2025}. This can result in only approximate locations of the eavesdroppers being detected in some scenarios, making precise physical layer security optimization through power allocation and trajectory design challenging. In such cases, as illustrated in Fig.~\ref{fig:typical_application}(d), satellites can control the UAVs to perform collaborative beamforming and provide a deployment scheme that minimizes the signal in all potential directions where eavesdroppers may be located in the low-altitude airspace. This ensures that physical layer security is achieved regardless of the orientation of the eavesdropper in LAWN applications.

\par In the following, we will introduce two case studies to detail the benefits of integrated satellite-UAV secure communications for real-world LAWN applications.

%
\section{Integrated Satellite-UAV Secure Relay Communications for LAWNs}
\label{sec:Case study 1}

\par In this section, we present a typical application of integrated satellite-UAV secure communications for LAWN scenarios. In network-limited regions of low-altitude airspace, where potential eavesdroppers may pose a threat to communications, satellites can control a collaborative UAV swarm to form a virtual antenna array and act as a relay to ensure physical layer security while completing the relay tasks~\cite{Sun2022}.

%
\subsection{Scenario and Scheme}
\label{ssec:case1_formualtion}

%
\begin{figure*}
  \centering
  \includegraphics[width=7 in]{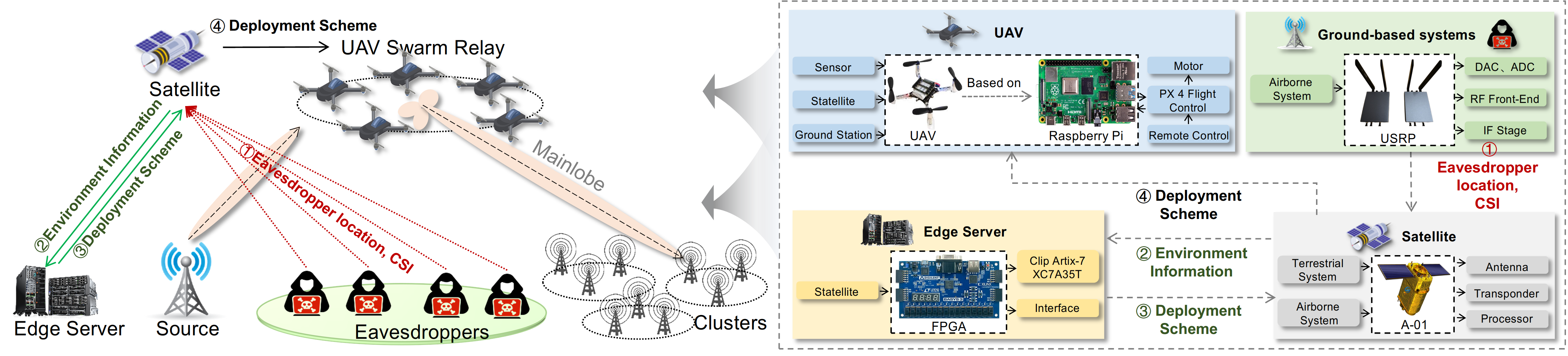}
  \caption{Integrated satellite-UAV secure relay communication system and the corresponding feasible testbed for LAWNs.}
  \label{fig:case1_scenario}
\end{figure*}

%
\begin{figure*}
  \centering
  \includegraphics[width=7in]{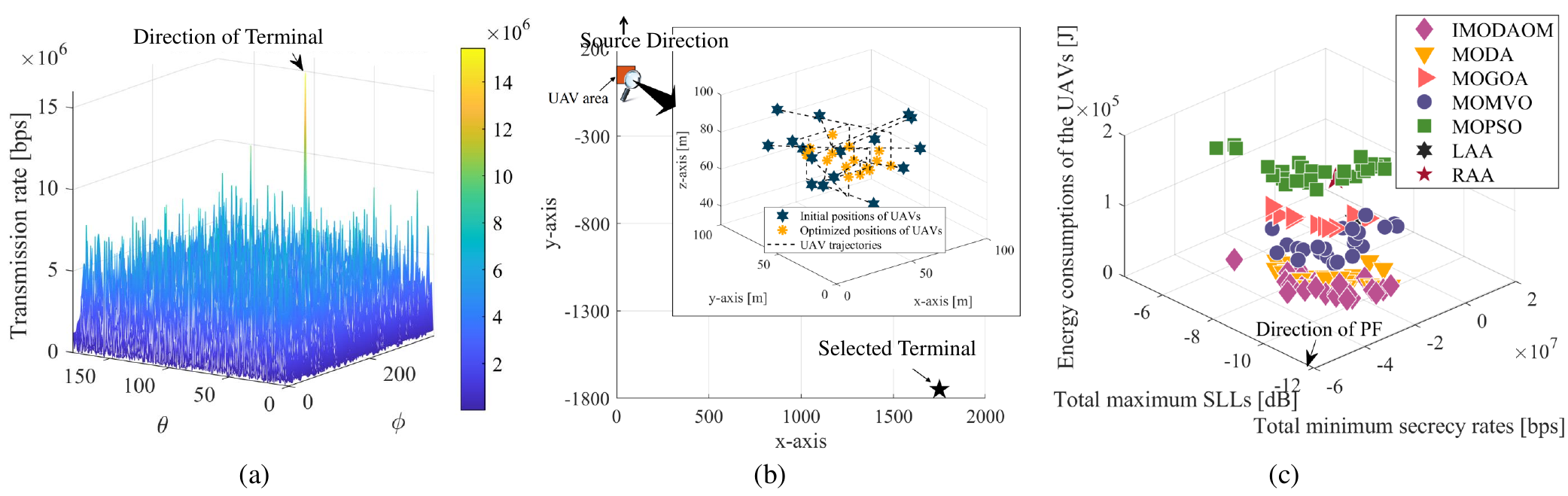}
  \caption{Simulation results: (a) Secrecy rate distributions; (b) Flight paths of UAVs; (c) Solution distributions.}
  \label{fig:case1_results}
\end{figure*}

\par As shown in Fig. \ref{fig:case1_scenario}, we investigate a terrestrial network system within the LAWN that needs relay assistance. Specifically, a source base station tries to connect with a few distant terrestrial terminals that may be supporting applications such as cargo tracking or aerial surveillance. We assume that the terminals can be separated into several clusters due to the obstruction among them. Note that terminals inside the same cluster can communicate effectively with each other, and direct connections between terminals within different clusters are impossible. Between the source base station and terminals, there are several eavesdroppers exist. We assume that some of the eavesdroppers can be detected by UAVs and satellites, while others are covert and cannot be detected. Our objective is to control the UAV swarm to act as an aerial relay and thereby transfer data from the source to the terminals in a secure way within the LAWN environment. 

\par The proposed scheme can be summarized as follows. First, the satellite acquires the UAV positions around the LAWN regions, while the UAVs and satellite perform remote sensing to find the eavesdropper locations. Then, the satellite transfers the information, including the positions of UAVs, terminals and eavesdroppers, and partial CSI, to the connected edge server. As such, the edge server runs the optimization method to obtain the UAV deployment strategy for the LAWN application. Finally, the satellite will send the deployment strategy to the UAV swarm and control the relay tasks. Compared to the direct relay of the satellite, this scheme can reduce the delay and enhance the throughput while saving the spectrum resources of the satellite in LAWN operations. 

\par The secure performance of this system is determined by the optimization method deployed on the edge server. Specifically, we aim to optimize the three aspects of the considered system. \textit{First}, we define the minimum secrecy rate as $R_{SEC}= [R_{R}-R_{E_{max}}]^+$, in which $R_{R}$ and $R_{E_{max}}$ are the transmission rates of a terminal and the known eavesdropper. Then, we maximize the total minimum secrecy rate when communicating with all clusters. \textit{Second}, we minimize the maximum sidelobe level of the virtual antenna array against unknown eavesdroppers. \textit{Third}, we reduce the total energy cost of UAVs by restricting their flying distance, an important consideration for sustainable LAWN operations. Note that the three optimization objectives are determined by the following decision variables: \textit{i}) The positions of these UAVs. \textit{ii}) The excitation current weights of these UAVs. \textit{iii}) The selection of receiver terminals in different clusters. \textit{iv}) The route of UAV swarm for transmitting data to different clusters. Note that this system can be tested by using the feasible testbed shown in Fig. \ref{fig:case1_scenario}. 

\par We employ multi-objective evolutionary algorithms to find near-optimal solutions with different objective trade-offs of the problem. Specifically, the multi-objective dragonfly algorithm (MODA) is enhanced to tackle this optimization problem. To make the MODA more suitable for solving the problem, we first use an orthogonal array to initial solutions for obtaining a high-performance distribution. Then, MODA cannot handle the sophisticated solution space, and thus we propose a multi-gravity hybrid solution update method based on partially mapped crossover to update the candidate solutions. By using these methods, a novel algorithm is proposed, namely, an improved MODA with orthogonal design-based solution initialization and multi-gravity hybrid solution update approaches (IMODAOM).

%
\subsection{Simulation Results}
\label{ssec:case1_results}

\par In our simulations for the LAWN application, we consider the scenario that there are four clusters, each containing eight terminals. Additionally, there is a 100 m $\times$ 100 m area in the middle of the entire field where 16 UAVs are dispersed. All terminals are kept at 3 km to 6 km away from the UAV area. Moreover, there are four known eavesdroppers and four unknown eavesdroppers in this system. The carrier frequency is selected at 900 MHz ISM band. In addition, the total noise power spectral density, pathloss exponent, bandwidth, and transmit power of each UAV are set as -155 dBm/Hz, 2.7, 20 MHz, and $0.1$ W, respectively. 

\par For comparison, we introduce two conventional virtual antenna arrays, in which the UAVs form uniform linear antenna arrays (LAA) and rectangular antenna arrays (RAA) with 0.5 m element spacing, respectively. Then, some peer multi-objective optimization approaches, including the multi-objective grasshopper optimization algorithm (MOGOA), the multi-objective multi-verse optimization (MOMVO), and the multi-objective particle swarm optimization (MOPSO), are introduced~\cite{Sun2022}.

\par The optimization results obtained by our IMODAOM and other peer methods are depicted in Fig. \ref{fig:case1_results}, showcasing the distributions of transmission rate, UAV deployment, and Pareto solution distributions of different algorithms for LAWN applications. As shown in Fig. \ref{fig:case1_results}(a), the transmission rate towards the target terminal is maximized compared to other directions. Furthermore, Fig. \ref{fig:case1_results}(b) reveals that the UAVs are suitably positioned, avoiding collisions or excessive dispersion which is critical for LAWN operations. Additionally, Fig. \ref{fig:case1_results}(c) illustrates the Pareto solution distributions of our IMODAOM and comparison algorithms. Note that each point denotes the objective values of the obtained solutions of the algorithms and different points show the various trade-offs of the solutions. As can be seen, our IMODAOM obtains solutions that are more wide-coverage and closer to the ideal Pareto front (PF). Thus, our IMODAOM effectively achieves the three objectives in the considered LAWN scenarios, rendering it a beneficial approach for secure low-altitude wireless communications.

%
\section{Integrated Satellite-UAV Secure Aerial Two-way Communications under Eavesdropper Collusion in LAWNs}
\label{sec:Case study 2}

\par In this section, we investigate a scenario where multiple eavesdroppers collude to achieve higher eavesdropping rates in the LAWN environment. In such cases, eavesdroppers can share their received signals to enhance their information-gathering capabilities. To tackle this security issue in low-altitude wireless networks, we propose a solution that involves two UAV swarms working in conjunction with a satellite, aimed at ensuring physical layer security even in the presence of eavesdropper collusion~\cite{Li2024}.

%
\begin{figure*}
  \centering
  \includegraphics[width=7 in]{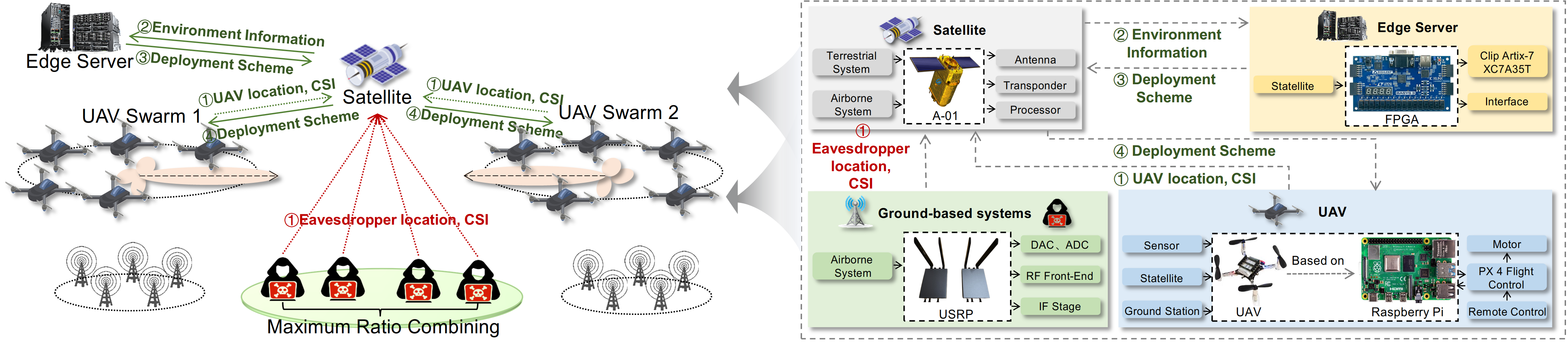}
  \caption{Secure aerial two-way communication system (under eavesdropper collusion) and the corresponding feasible testbed in LAWNs enabled by integrated satellite-UAV.}
  \label{fig:case2_scenario}
\end{figure*}

%
\begin{figure*}
  \centering
  \includegraphics[width=7 in]{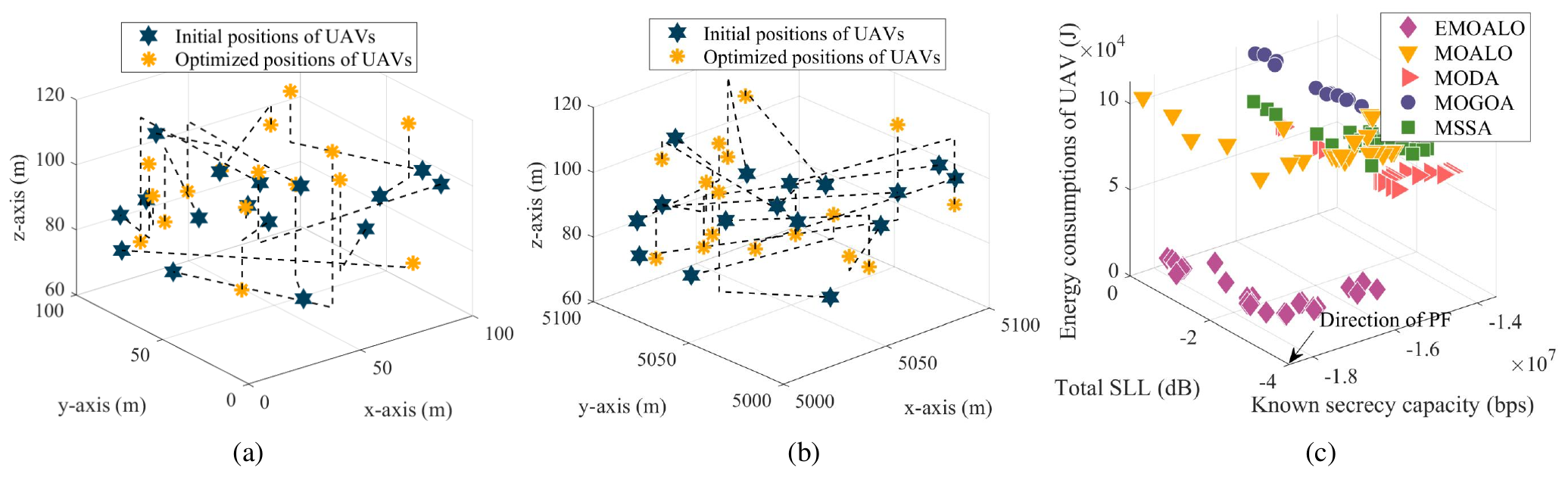}
  \caption{Simulation results: (a) Flight paths of UAVs in UAV swarm 1; (b) Flight paths of UAVs in UAV swarm 2; (c) Solution distribution.}
  \label{fig:case2_results}
\end{figure*}

%
\subsection{Scenario and Scheme}
\label{ssec:case2_formualtion}

\par In the scenario depicted in Fig. \ref{fig:case2_scenario}, we consider a two-way aerial communication system in the LAWN where two UAV swarms need to exchange large amounts of data, such as might be required for coordinated cargo delivery or air taxi services. Note that the two UAV swarms are located far apart from each other in the low-altitude airspace and there is no available infrastructure for direct communication. Moreover, there are several eavesdroppers randomly placed in the area, some of which can be detected by UAVs or satellites, while others remain undetected. These eavesdroppers collude via a maximum ratio combining approach, posing a severe security challenge to LAWN operations. In this case, using a satellite to direct relay between UAV swarms leads to increased latency and higher bandwidth occupation. As such, our objective is to control the two UAV swarms to form two virtual antenna arrays, enabling secure data exchange in the low-altitude wireless network.

\par The scheme is the same as the case study in Section \ref{sec:Case study 1}. First, the satellite obtains the UAV positions of two UAV swarms, while the UAVs and the satellite perform remote sensing to find the eavesdropper locations in the LAWN environment. Then, the satellite transfers the information, including the positions of UAVs and eavesdroppers, and partial CSI, to the connected edge server. As such, the edge server runs the optimization method to obtain the UAV deployment strategy for the low-altitude application. Finally, the satellite sends the deployment strategy to the two UAV swarms, and then two UAV swarms perform two-way communication in the low-altitude wireless network. 

\par Likewise, the sum secrecy rates of two UAV swarms are determined by the optimization method deployed on the edge server. Since we cannot obtain the exact information of unknown eavesdroppers in the LAWN, we aim to take two measures simultaneously to optimize the achievable secrecy capacity. \textit{First}, we can define the known secrecy capacity as $R_{SEC}= [R_{C}-R_{E_{MRC}}]^+$, where $R_{C}$ and $R_{E_{MRC}}$ are the transmission rates of the counterpart UAV swarm and the colluded eavesdroppers. Following this, we maximize the known secrecy capacity in two-way communications to reduce the signals obtained by the known eavesdroppers. \textit{Second}, we minimize all the signals except the target direction, thereby avoiding potential unknown eavesdroppers in the low-altitude airspace. \textit{Third}, we also reduce the energy costs of the two UAV swarms consumed in fine-tuning UAVs' positions, which is particularly important for extending endurance in LAWN applications. To achieve these objectives, we can jointly control these decision variables: \textit{i}) The positions of UAVs in two UAV swarms. \textit{ii}) The excitation current weights of UAVs in two UAV swarms. \textit{iii}) The selection of receiver UAVs of two UAV swarms. Also, this system can be tested by using the testbed shown in Fig. \ref{fig:case2_scenario}. 

\par This optimization problem is an NP-hard problem with multiple optimization objectives and large-scale mixed decision variables. Thus, we also use and enhance a multi-objective evolutionary algorithm, \textit{i.e.}, multi-objective ant-lion optimizer (MOALO), to solve this optimization problem for LAWN applications. \textit{First}, we use random walk based on the initial positions of the UAVs to generate an initial population. \textit{Second}, we improve the population evolution in the early stage of iteration and improve the population diversity in the latter stage of iteration. \textit{Third}, we introduce an integer mutation method to update the candidate solutions. By using these improved factors, our proposed enhanced MOALO (EMOALO) may have a better solving ability for the formulated problem in low-altitude wireless networks.

%
\subsection{Simulation Results}
\label{ssec:case2_results}

\par To demonstrate the effectiveness of EMOALO for LAWN applications, we consider a scenario where the UAV areas are all set as 100 m $\times$ 100 m in low-altitude airspace. Likewise, the carrier frequency, transmit power of each UAV, and path loss coefficient are set as 2.4 GHz, 0.1 W, and 2, respectively. Then, we introduce MODA, MOMVO, multi-objective salp swarm algorithm (MSSA), and conventional MOALO as comparison algorithms~\cite{Li2024}. We show two UAV swarms' position changes in Figs. \ref{fig:case2_results}(a) and \ref{fig:case2_results}(b). As can be seen, only fine-tuning UAVs' position can achieve physical layer security communications in the LAWN environment. Additionally, Fig. \ref{fig:case2_results}(a) shows the Pareto solution distributions of the aforementioned multi-objective optimization methods. As can be seen, the Pareto solutions obtained by our EMOMVO are closer to the ideal PF, indicating superior performance compared to other algorithms. These results highlight the effectiveness of our EMOALO in achieving high-quality solutions for the considered LAWN scenario in low-altitude wireless networks.

%
\section{Practicality Analysis for Lawn Security Implementation}

\par We conduct a practicality analysis for our proposed methodology and encryption/decryption techniques in LAWN environments. As shown in Figs.~\ref{fig:case1_scenario} and~\ref{fig:case2_scenario}, our implementation utilizes the Raspberry Pi 4B as the UAV flight control system, representing a configuration widely adopted in prevalent UAV platforms (\textit{e.g.}, PX4 autopilot). We operate under the assumption that a UAV swarm executes a parallel distributed version of the swarm intelligence algorithm, with the optimization objective calculation step omitted since this is typically substituted with proxy models in operational deployments~\cite{Li2024}. For comparative analysis, we incorporate three conventional encryption/decryption protocols: data encryption standard (DES), advanced encryption standard (AES), and Rivest-Shamir-Adleman (RSA).

\par Experimental findings demonstrate that our proposed algorithm completes a single calculation iteration within about 40 seconds. Furthermore, the processing times for encrypting and decrypting 200 MB of data using DES, AES, and RSA are measured at 12.07 s, 9.29 s, and 1567.59 s, respectively. Consequently, when data transmission exceeds approximately 1 GB, our proposed methodology demonstrates significant computational efficiency advantages. This performance differential stems from the fact that traditional encryption/decryption methods require continuous calculation throughout the data transfer process, whereas our approach necessitates only a one-time calculation. Therefore, the computational benefits of our methodology become increasingly pronounced as data volume expands, making it particularly suitable for bandwidth-intensive LAWN applications.

%
\section{Challenges and Future Directions for LAWN Security}
\label{sec: challenges_and_future_direction}

\par The aforementioned typical applications and case studies offer some ideas and insights into integrated satellite-UAV secure networks for LAWN applications. To further promote the application of this approach in low-altitude wireless networks, several future directions can be pursued.

\par \textbf{Detection of Eavesdroppers in LAWNs:} The remote sensing capabilities of satellites, as well as the cameras and radars equipped on UAVs, can aid in detecting the approximate locations of potential eavesdroppers in low-altitude airspace~\cite{Yan2025}. However, precise localization of eavesdroppers is crucial for achieving effective physical layer security in LAWN applications. Therefore, exploring how to synergistically leverage the multi-type sensing abilities of satellites and UAVs to accurately identify the precise positions of eavesdroppers in the dynamic LAWN environment holds promise as a future research direction.

\par \textbf{Protocol Design for LAWN Applications:} The dynamic and unpredictable nature of integrated satellite-UAV networks in low-altitude airspace presents unique challenges for protocol design, as traditional network/transport protocols may not perform optimally in such environments due to rapidly changing topology information, high communication overhead for refreshing outdated information, and sudden drops in link quality resulting in reduced transmission rates and decreased resource utilization. To tackle these challenges, innovative protocol design approaches specific to LAWN applications are needed, possibly incorporating AI-based approaches for adaptive protocol selection and parameter tuning.

\par \textbf{Advanced Optimization Techniques for LAWNs}: The use of advanced optimization techniques, such as machine learning-based algorithms, hybrid optimization algorithms, or meta-heuristic algorithms, could be further explored to enhance the performance of UAV collaborative secure communication systems in the low-altitude wireless networks. These techniques could help to optimize the system parameters, UAV trajectories, and resource allocations more efficiently and effectively, leading to improved communication performance and security for diverse LAWN applications.

\par \textbf{Synchronization in LAWNs:} Integrated satellite-UAV systems in low-altitude environments are characterized by fast-moving UAV swarms that may be widely distributed, resulting in significant differential propagation delays between UAVs at the edge and those at the center, particularly at low elevation angles. This can lead to synchronization issues, as collaborative beamforming requires UAVs to be synchronized in terms of time, phase, and frequency, so that all antennas can send signals to the receiver simultaneously. Addressing these synchronization challenges is crucial for achieving effective collaborative beamforming in integrated satellite-UAV systems supporting LAWN operations.

\par \textbf{Testbed Implementation and Standardization for LAWNs:} To facilitate the practical deployment of secure low-altitude wireless networks, comprehensive testbeds need to be designed and implemented. These testbeds should realistically model the challenges of LAWN environments, including variable weather conditions, airspace restrictions, and dynamic mobility patterns. Furthermore, standardization efforts are essential to ensure interoperability between different UAV platforms, satellite systems, and ground infrastructure, enabling seamless integration of these components in real-world LAWN applications while maintaining robust security features.

\par \textbf{Satellite Collaborative Beamforming for LAWN Support:} Similar to UAVs, collaborative beamforming can also be employed by multiple satellites to enhance physical layer security in LAWNs. Satellites, being immune to wind and jitter, offer the advantage of more stable communication links for supporting low-altitude operations. However, due to the relatively fixed orbits of satellites, there are challenges in designing effective satellite collaborative beamforming schemes that account for their orbital constraints while optimizing support for dynamic LAWN applications.

%
\section{Conclusion}
\label{sec:conclusion}

\par This paper has investigated collaborative beamforming-based physical layer security schemes achieved by integrated satellite-UAV systems for low-altitude wireless networks. We have begun by providing an overview of integrated satellite-UAV networks, physical layer security, and collaborative beamforming in LAWNs. Subsequently, we have highlighted some opportunities for integrated satellite-UAV secure communications applications in LAWN environments, such as physical layer security in data dissemination, data relay, scenarios involving eavesdropper collusion, and imperfect eavesdropper information. Additionally, we have presented two case studies and a practicality analysis to demonstrate the effectiveness of the proposed systems in realistic LAWN scenarios. Finally, we have identified current issues and suggested future directions for further research.


%



\ifCLASSOPTIONcompsoc
\else
\fi


\ifCLASSOPTIONcaptionsoff
  \newpage
\fi



%

\bibliographystyle{IEEEtran}
\bibliography{myref}

\end{document}